\documentclass[a4paper, 10pt, twocolumn]{article}

\usepackage{amsfonts,amssymb,amsmath,amsthm}
\usepackage{graphicx}
\usepackage[footnotesize]{caption}
\usepackage{subfig}
\usepackage{microtype}

\usepackage{enumitem}
\setlist[itemize]{noitemsep, topsep=0pt}

\usepackage[top=1.5cm, left=1.5cm, right=1.5cm, bottom=1.5cm]{geometry}

\usepackage{algorithm}
\usepackage{algorithmicx}
\usepackage[noend]{algpseudocode}

\algblockdefx{MRepeat}{EndRepeat}{\textbf{repeat}}{}
\algnotext{EndRepeat}
\usepackage[export]{adjustbox}

\usepackage{tabularx}
\usepackage{etoolbox}

\algrenewcommand\algorithmicindent{0.6em}%

\newcommand{\dom}{\mathcal{D}}

\usepackage{tikz}

\usetikzlibrary{tikzmark}
\usetikzlibrary{calc}

\usepackage{pgfplots}
\pgfplotsset{compat=1.8}
\usepackage{pgfplotstable}
\usepgfplotslibrary{groupplots}

\errorcontextlines\maxdimen

\newcommand{\ALGtikzmarkcolor}{black}
\newcommand{\ALGtikzmarkextraindent}{4pt}
\newcommand{\ALGtikzmarkverticaloffsetstart}{-.5ex}
\newcommand{\ALGtikzmarkverticaloffsetend}{-.5ex}
\makeatletter
\newcounter{ALG@tikzmark@tempcnta}

\newcommand\ALG@tikzmark@start{%
    \global\let\ALG@tikzmark@last\ALG@tikzmark@starttext%
    \expandafter\edef\csname ALG@tikzmark@\theALG@nested\endcsname{\theALG@tikzmark@tempcnta}%
    \tikzmark{ALG@tikzmark@start@\csname ALG@tikzmark@\theALG@nested\endcsname}%
    \addtocounter{ALG@tikzmark@tempcnta}{1}%
}

\def\ALG@tikzmark@starttext{start}
\newcommand\ALG@tikzmark@end{%
    \ifx\ALG@tikzmark@last\ALG@tikzmark@starttext
    \else
        \tikzmark{ALG@tikzmark@end@\csname ALG@tikzmark@\theALG@nested\endcsname}%
        \tikz[overlay,remember picture] \draw[\ALGtikzmarkcolor] let \p{S}=($(pic cs:ALG@tikzmark@start@\csname ALG@tikzmark@\theALG@nested\endcsname)+(\ALGtikzmarkextraindent,\ALGtikzmarkverticaloffsetstart)$), \p{E}=($(pic cs:ALG@tikzmark@end@\csname ALG@tikzmark@\theALG@nested\endcsname)+(\ALGtikzmarkextraindent,\ALGtikzmarkverticaloffsetend)$) in (\x{S},\y{S})--(\x{S},\y{E});%
    \fi
    \gdef\ALG@tikzmark@last{end}%
}

\usepackage{etoolbox}

\apptocmd{\ALG@beginblock}{\ALG@tikzmark@start}{}{\errmessage{failed to patch}}
\pretocmd{\ALG@endblock}{\ALG@tikzmark@end}{}{\errmessage{failed to patch}}
\makeatother

\makeatletter
\newcommand{\multiline}[1]{%
  \begin{tabularx}{\dimexpr\linewidth-\ALG@thistlm}[t]{@{}X@{}}
    #1
  \end{tabularx}
}
\makeatother

\renewcommand{\H}{\mathbf{H}}

\newcommand{\Gc}{\mathcal{G}}

\newcommand{\bmu}{\mu}
\newcommand{\bmeta}{\eta}
\newcommand{\bPhi}{\boldsymbol{\Phi}}
\newcommand{\bt}{\boldsymbol{\theta}}

\newcommand{\bw}{\boldsymbol{w}}

\newcommand{\w}{\boldsymbol{w}}
\newcommand{\m}{\boldsymbol{m}}
\newcommand{\x}{\mathbf{x}}
\newcommand{\y}{\mathbf{y}}
\newcommand{\mwt}{\bmu_{\w,\bt}}

\newcommand{\eqdef}{\overset{\mathrm{def.}}{=}}

\newcommand{\s}{\boldsymbol{s}}

\title{Boosting the Sliding Frank-Wolfe solver for 3D deconvolution}

\author{Jean-Baptiste Courbot, Bruno Colicchio\\
\footnotesize IRIMAS UR 7499, Universit\'{e} de Haute-Alsace, Mulhouse, France
}
\date{\empty} 

\renewenvironment{abstract}{\bf\small {\em\ Abstract---}}{}

\begin{document}

\maketitle

\begin{abstract}
In the context of gridless sparse optimization, the Sliding Frank Wolfe algorithm recently introduced has shown interesting analytical and practical properties. 
Nevertheless, is application to large data, such as in the case of 3D  deconvolution, is computationally heavy. 
In this paper, we investigate a strategy for leveraging this burden, in order to make this method more tractable for 3D deconvolution.
We show that a boosted SFW can achieve the same results in a significantly reduced amount of time. 
\end{abstract}

\vspace{-.5em}%
\section{Introduction}
\label{sec:introduction}

\subsection{Observation model}

In this paper, we consider the problem of 3D deconvolution of some volume $\y$ containing a small number of atoms. 
We assume that $\y$ results from some measure $\bmu_{\w,\bt}$ observed through an imaging operator $\bPhi$:
\begin{equation}
\y =\bPhi \bmu_{\w,\bt} + \boldsymbol{\epsilon}
\label{eq:obs_model1}
\end{equation}
Here  and in the following, we consider for the measures $\mwt $ a positive weighted Dirac mass sum of the form
{$\sum_{n=1}^{N} w_n \delta_{\bt_n}$},
with the weight vector $\bw = \{w_1, \ldots, w_N \} \in \mathbb{R}_+^N$.
$\bt_n$ locates, for each atom, its parameters within the bounded domain $\dom$.
Without loss of generality, we choose to use generalized isotropic Gaussians as atoms, such that  $\forall 1 \leq n \leq N$ and $\forall \s \in \mathbb{R}^3$:
\begin{equation}
\Gc(\bt_n, w_n ; \s ) = w _n\exp\left(-  \frac{1}{2\sigma_n^{d_n}} \| \m_n - \s \|^{d_n}  \right)
\end{equation}
so that $\bt_n = \{ \m_n, \sigma_n, d_n  \} \in \dom \subset \mathbb{R}^3 \times \mathbb{R}_+^2$.

Furthermore, $\bPhi$ embeds a point spread function (PSF) $\H$ that blurs the observations of atoms. Rephrasing~\eqref{eq:obs_model1}, we have:
\begin{equation}
\y = \H \ast  \sum_{n=1}^{N} \Gc(\bt_n, w_n) + \boldsymbol{\epsilon}
\label{eq:obs_model}
\end{equation}
The problem handled here consists in estimating $N$,  and  $\{\bt_n, \bw_n\}_{n=1}^{n=N}$ while knowing only $\y$ and $\H$.

\subsection{Continuous sparsity and the SFW algorithm}

\begin{algorithm}[!h]\caption{Sliding Frank-Wolfe \cite{denoyelle2018sliding}\label{alg:sfw}}
	\begin{algorithmic}
		\Require $\y$, PSF $\H$, $\lambda$
		\Ensure Estimated minimizer $\hat{\bmu}_{\bw,\bt}$ of~\eqref{eq:norm1} 
		\State Initialization: $\bmu_{\bw^{(0)},\bt^{(0)}} = {0}$.
\MRepeat\,(step $k$):
\State \multiline{1. Compute  $\max_{\bt \in \dom}\bmeta^{(k)}$ by local ascent using \textit{e.g.} BFGS, starting from a maximum attained on a grid.}
\If{$\max_{\bt \in \dom}\bmeta^{(k)} > 1$}
\State 2. Augment the support: $	\bt^{(k)}\hspace{-.25em}=\hspace{-.19em}\bt^{(k-1)} \cup {\arg\max_{\bt \in \dom}} ~\bmeta^{(k)}$
\State {3. Adjust weights only (LASSO): \\
\qquad\qquad  \qquad \qquad \qquad  $\tilde{\bw}^{(k)} = \underset{\w \in \mathbb{R}_+^{k}}{\arg \min}~
		C(\y,\bmu_{\w,\bt^{(k)}},\lambda)$}%
		\State \multiline{4. Local {descent} on all parameters using \textit{e.g.} BFGS, starting at $\bmu_{\tilde{\w}^{(k)},\bt^{(k)}}$:}\\
\qquad\qquad   \qquad $	{\bw}^{(k)},\bt^{(k)} = \underset{
			\w \in \mathbb{R}_+^{k},\bt \in \dom^{k}
		}{\text{local descent of}}~C(\y,\bmu_{\w,\bt},\lambda)$
		\State 5. Remove zero-weighted masses, update the measure: \\%
\qquad\qquad  \qquad \qquad \qquad$\bmu_{\bw^{(k)},\bt^{(k)}} = \sum_{n=1}^k w_n^{(k)} \delta_{\bt_n^{(k)}}$%
\Else
\State ~\textbf{end} of SFW
\EndIf
\EndRepeat
	\end{algorithmic}
\end{algorithm}
The problem handled here is stated in a sparse continuous fashion: there are a few atoms to search for, and they do not lie on a pre-established grid or dictionary. 
Hence, it is desirable to work in a fully continuous setting~\cite{bredies2013inverse,candes2014towards,de2012exact}, replacing the sparsity-promoting $\ell_1$ norm with its continuous counterpart, the total variation of measures.

To solve the problem, coined BLASSO by~\cite{azais2015spike}, several approaches exist.  In \cite{candes2014towards,tang2013compressed}, the problem is recast as a semi-definite program, whereas the ADCG solver proposed in \cite{boyd2017alternating,bredies2013inverse} relies on an alternating gradient based method which progressively adds Dirac masses. Recently, a variant of the ADCG called Sliding Frank-Wolfe (SFW) appeared in \cite{denoyelle2018sliding}, which is guaranteed to converge in a finite number of steps under suitable assumptions.

In this paper, we focus on the SFW algorithm and its application to the inversion of~\eqref{eq:obs_model}.
	To retrieve $\mwt$ in a sparse fashion, we minimize among all non-negative measures $\mu$ the following criterion:
	\begin{equation}
	C(\y,	\mwt,\lambda) = \frac{1}{2}\left\Vert \y - \bPhi\mwt \right\Vert_2^2 + \lambda  \mwt(\dom)
	\label{eq:norm1}
	\end{equation}
	where $\lambda > 0$ is the regularization parameter, and $\mu(\dom)$ denotes the total mass of the non-negative measure $\mu$. 
{In the case of a sum of Dirac masses, $\mwt(\dom)=\sum_{n=1}^{N} w_n$.} 

SFW is a greedy algorithm, which adds iteratively Dirac masses to the estimated measure $\mu$. To do so, SFW is ruled by a certificate $\eta: \dom \rightarrow \mathbb{R}$ indicating where to append masses, and when to stop the algorithm. It is defined, at step $k$, as:
\begin{equation}
\bmeta^{(k)} \eqdef \frac{1}{\lambda} \bPhi^\top \left(\y - \bPhi\bmu_{\bw^{(k-1)},\bt^{(k-1)}} \right)
\label{eq:eta}
\end{equation}
Informally, it can be  seen as the result of the convolution between the residual and the volume of an atom located at one point in $\dom$.  
Algorithm~\ref{alg:sfw} depicts SFW, and further details on this algorithm can be found in ~\cite{denoyelle2018sliding}.
%

\vspace{-.5em}%
\section{Boosting the SFW solver}

When applying the SFW solver to invert~\eqref{eq:obs_model}, several observations can be made:
\begin{itemize}
\item in most cases SFW is efficient and retrieves the $K$ atoms provided in exactly $K$ steps. This has already been noted in~\cite{denoyelle2018sliding}, in which the convergence in a finite number of steps has been proven under mild assumptions.
\item most computation time is spent in the local descents of step 4, which handle all $k \times 6$ parameters (6 being, in our case, the dimension of $\dom$). This seems computationally wasteful: in all iterations except the last, the fine-tuned parameters are modified afterwards.
\item this step only marginally decreases $C$ as defined in~\eqref{eq:norm1}. We noted that this is related to the precision of the result of step~1: when the local ascent maximizing $\eta$ yields a relevant result, the local descent of step 4 does not decrease $C$.
\end{itemize}
Based on these observations, we propose a boosted version of the SFW algorithm, denoted BSFW, which removes all but the last local descent. 
As in SFW, the certificate $\eta$ locates new atoms and indicates when the algorithm should stop. Then, at iteration $k$, either:
\begin{itemize}
\item $\max_{\bt \in \dom}\bmeta^{(k)} > 1$, so a new Dirac mass is added, and $\bw$ is adjusted,
\item $\max_{\bt \in \dom}\bmeta^{(k)} < 1$, so a local descent is made to adjust $(\bw, \bt)$, and then BSFW stops.
\end{itemize}
The procedure is summarized in Algorithm~\ref{alg:bsfw}.

\begin{algorithm}\caption{Boosted Sliding Frank-Wolfe \label{alg:bsfw}}
	\begin{algorithmic}
		\Require $\y$, PSF $\H$, $\lambda$
		\Ensure Estimated minimizer $\hat{\bmu}_{\bw,\bt}$ of~\eqref{eq:norm1} 
		\State Initialization: $\bmu_{\bw^{(0)},\bt^{(0)}} = {0}$.
\MRepeat:
\State \multiline{1. Compute  $\max_{\bt \in \dom}\bmeta^{(k)}$ by local ascent using \textit{e.g.} BFGS, starting from a maximum attained on a grid.}
\If{$\max_{\bt \in \dom}\bmeta^{(k)} > 1$}
\State 2. Augment the support: $	\bt^{(k)}\hspace{-.25em}=\hspace{-.19em}\bt^{(k-1)} \cup {\arg\max_{\bt \in \dom}} ~\bmeta^{(k)}$
\State {3. Adjust weights only (LASSO): \\
\qquad\qquad  \qquad \qquad \qquad  $\tilde{\bw}^{(k)} = \underset{\w \in \mathbb{R}_+^{k}}{\arg \min}~
		C(\y,\bmu_{\w,\bt^{(k)}},\lambda)$}%
		\State 4. Remove zero-weighted masses, update the measure: \\%
\qquad\qquad  \qquad \qquad \qquad$\bmu_{\bw^{(k)},\bt^{(k)}} = \sum_{n=1}^k \tilde{w}_n^{(k)} \delta_{\bt_n^{(k)}}$%
\Else 
	\State \multiline{2b. Local {descent} on all parameters using \textit{e.g.} BFGS, starting at $\bmu_{\tilde{\w}^{(k)},\bt^{(k)}}$:}\\
\qquad\qquad   \qquad $	{\bw}^{(k)},\bt^{(k)} = \underset{
			\w \in \mathbb{R}_+^{k},\bt \in \dom^{k}
		}{\text{local descent of}}~C(\y,\bmu_{\w,\bt},\lambda)$
	\State 3b. Remove zero-weighted masses, update the measure: \\%
\begin{flushright}$\bmu_{\bw^{(k)},\bt^{(k)}} = \sum_{n=1}^k w_n^{(k)} \delta_{\bt_n^{(k)}}$%
\end{flushright}%

		\State \textbf{end} of BSFW
\EndIf
\EndRepeat
	\end{algorithmic}
\end{algorithm}

Note that steps 1--4 of the BSFW algorithm form a Frank-Wolfe algorithm, so BSFW has similar convergences properties. 
However, the BSFW does not benefits from the nicer convergence properties of SFW (explained in~\cite{denoyelle2018sliding}), because local descents were removed.

In addition to the new algorithm presented above, several implementation points have been leveraged to improve the speed of both SFW and BSFW:
\begin{itemize}
\item convolutions are made in Fourier domain,
\item $\nabla C$ is computed analytically, so as to avoid a costly numerical approximation,
\item the computation of $C$ and $\nabla C$ is parallelized across atoms,
\item a lookup table for convolution in $\dom$ is computed once in order to help the computations of $\eta$ at step 1.
\end{itemize}

\vspace{-.5em}%
\section{Numerical results}

\begin{figure}
\,\hfill\includegraphics[width=0.25\linewidth,frame]{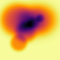}\hfill 
\includegraphics[width=0.25\linewidth,frame]{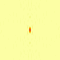}\hfill
\includegraphics[width=0.25\linewidth,frame]{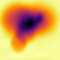}\hfill\,

\caption{2D slices of 3D volumes corresponding to $\sum_{n=1}^{N} \Gc(\bt^*_n, w^*_n)$ (left), $\H$ (middle) and $\y$ (right). Note that $\H$ is invariant by rotation around the vertical axis. \label{fig:ex}}
\end{figure}

In this section, we compare the numerical behavior of SFW and BSFW. 
We use a PSF $\H$ appearing in tomographic diffractive microscopy~\cite{simon2019tomographic}, which significantly blur along the optical axis. Figure~\ref{fig:ex} exemplifies the volume handled in our problem.

When inverting~\ref{eq:obs_model} with SFW and BSFW, we are mainly interested in (a) the computation time and (b), the value of $C$, in order to measure the loss of precision (or no) from SFW to BSFW.

\begin{figure}
\centering
\input{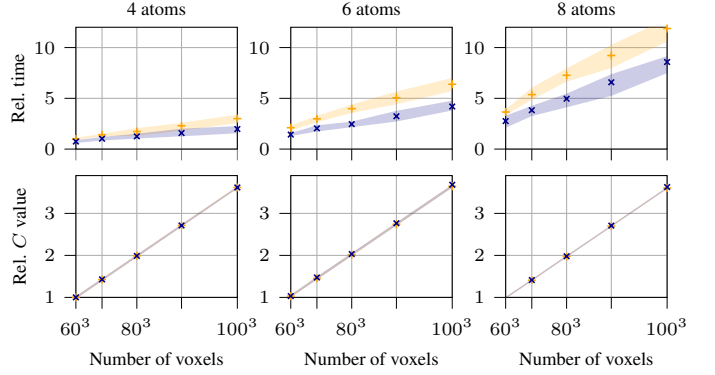}
\vspace{-1.5em}
\caption{Averaged numerical results for SFW and BSFW (marks), with the first and third quartiles of the results depicted within the colored regions. The first line depicts the computation time, relatively to the smaller  value encountered by  SFW ($60^3$ voxels and 4 atoms). The second line shows, similarly,  the relative values of $C$~\eqref{eq:norm1} at the end of SFW and BSFW.\label{fig:res} }
\end{figure}

Both algorithms were studied while varying the size of the considered volumes (from $60^3$ to $100^3$ voxels) and the number of Dirac mass to retrieve (4, 6 and 8) using a fixed $\lambda = 0.2$ and a Gaussian white noise.
Each experiment was repeated 30 times with different values of the real $\mu^*$, and Fig.~\ref{fig:res} presents the averaged results. 

\noindent These results can be summarized as follows:
\begin{itemize}
\item BSFW attains almost exactly the same criterion value as SFW in all the studied cases. 
\item BSFW improves notably the computation times: on average it runs $31\%$ faster than FSW for the same task.  
\end{itemize}
However the improvements are not as important as expected from the removal of almost all the costly local descents. After inspecting the results, we observed that in some cases, repeating steps 1--4 yields more Dirac masses then necessary, which are costly to manage within the local descents of step 2b. These additional masses are only removed at step 3b, so the final results does yields the adequate number of atoms.

\vspace{-.5em}%
\section{Discussion}
\label{sec:conclusion}

We studied the potential speed improvements of the SFW algorithm, by removing costly intermediate local descent steps. We showed that this alternative does run significantly faster than SFW, at the cost of following a non-optimal optimization path and thus preventing BSFW to run even faster.

Several perspectives are considered as future works, among which the improvement of the optimization path within BSFW, and the joint search for an adequate regularization parameter.  


\clearpage

\bibliographystyle{plain}
\bibliography{biblio.bib}
%
%

\end{document}